\documentclass[sigconf]{acmart}

\usepackage{subfigure}

\AtBeginDocument{%
  }

\renewcommand\footnotetextcopyrightpermission[1]{} 
\setcopyright{none}
\acmConference[WWW '25]{April 28--May 02, 2025}{April 28--May 02, 2025}{Sydney, Australia}

\acmISBN{978-1-4503-XXXX-X/18/06}

\begin{document}

\title[Evaluating Next.js: A Comparative Analysis with React.js]{Evaluating the Efficacy of Next.js: A Comparative Analysis with React.js on Performance, SEO, and Global Network Equity}

\author{Swostik Pati}
\email{swostik.pati@nyu.edu}
\affiliation{%
  \institution{New York University Abu Dhabi}
  \country{United Arab Emirates}
}

\author{Yasir Zaki}
\email{yasir.zaki@nyu.edu}
\affiliation{%
  \institution{New York University Abu Dhabi}
  \country{United Arab Emirates}
}

\renewcommand{\shortauthors}{Pati et al.}

\begin{abstract}
  This paper investigates the efficacy of Next.js as a framework addressing the challenges posed by React.js, particularly in performance, SEO, and equitable web accessibility. By constructing identical websites and web applications in both frameworks, we aim to evaluate the frameworks' behavior under diverse network conditions and capabilities. Beyond quantitative metrics like First Contentful Paint (FCP) and Time to Interactive (TTI), we incorporate qualitative user feedback to assess real-world usability. Our motivation stems from bridging the digital divide exacerbated by client-side rendering (CSR) frameworks and validating investments in modern technologies for businesses and institutions. Employing a novel LLM-assisted migration workflow, this paper also demonstrates the ease with which developers can transition from React.js to Next.js. Our results highlight Next.js's promise of better overall performance, without any degradation in user interaction experience, showcasing its potential to mitigate disparities in web accessibility and foster global network equity, thus highlighting Next.js as a compelling framework for the future of an inclusive web.
\end{abstract}




\maketitle

\section{Introduction}
In the dynamic realm of web development, the transition from traditional server-side rendering (SSR) to client-side rendering (CSR) has marked a significant evolution in how web applications are designed and interacted with. SSR operates by having the webpage's content be first generated on the server and then sent over the network to the user's browser in its final HTML form. This method, while stable and reliable, often results in slower response times to user interaction, as each request necessitates a new page to be served from the server. To counter this, CSR was introduced as a more modern approach that shifted the rendering operations to the browser. Web applications download a minimal HTML page from the server, which then uses loads of JavaScript files to generate the rest of the page's content dynamically. This shift not only reduces the server load but also significantly enhances the interactivity and fluidity of user interfaces. Web apps became more responsive, providing a smoother user experience as the browser could re-render only specific parts of the page in response to user actions without needing to fetch a new page from the server each time. 

However, as the landscape evolved, it became evident that CSR, despite its improvements in reactivity and user interface fluidity, has introduced significant challenges. These include suboptimal search engine optimization (SEO), poorer performance~\cite{Iskandar2020CSRvsSSR}, and an increased demand on the client resources. As such, it led on to a disparate web experience across varied network conditions, highlighting a digital divide that is more pronounced in low-bandwidth environments. One major example of this is the case of React.js~\cite{ReactWebsite}, which exemplifies both the strengths and weaknesses of CSR. React.js is a client-side JavaScript framework that was widely adopted around the world ever since its launch and is currently the second most used web framework with almost 40\% market share~\cite{Statista2024WebFrameworks}. However, even with such wide adoption rates, frequent updates, and several workarounds and optimizations, React.js still suffers from all the aforementioned issues and challenges of CSR.

All of the above has led to an increased interest in what are known as hybrid frameworks, such as Next.js~\cite{NextjsWebsite} (a JS framework built over React), which mitigates these issues by combining the best features of SSR and CSR. Next.js promises to enhance SEO and performance through techniques like pre-fetching, incremental static regeneration, and static site generation. In addition, through its default caching mechanisms and use of server components (which would reduce the client-side overhead), it effectively optimizes resource utilization, ensuring that users worldwide, regardless of their network, receive a faster and more seamless web experience. 

Due to the benefits of Next.js, the web development world has been fast to adopt this framework, with Next.js having become the 4th most used web framework globally (i.e., 18\% market share~\cite{Statista2024WebFrameworks}) in less than ten years since its inception.  Given the widespread migration towards Next.js, this study seeks to preemptively address potential pitfalls by evaluating Next.js and benchmarking its performance against its parent library, React.js, across various performance metrics (i.e., loading times, lighthouse scores), network conditions, and device processing capabilities. By doing so, we set out to understand whether Next.js can truly live up to its promises. 

In this paper we plan to answer the following research questions: \textbf{RQ1} Does Next.js have an overall faster performance compared to React.js across different network conditions and device specifications? \textbf{RQ2} Does Next.js have better Search Engine Optimization (SEO) metrics than React.js? \textbf{RQ3} Beyond the quantitative evaluations, does Next.js match---if not exceed---React.js in terms of user interaction experience?

\section{Motivation}

The motivation behind this work is twofold: ensuring that the world as a whole gets to experience the web in a similar way, bridging the digital divide that is currently present due to disparities across network bandwidths and device processing capabilities~\cite{digitaldivide}; and ensuring that when content providers migrate into newer web technologies such as Next.js, they are actually worth the investment. 

\subsection{Towards Global Network equity}

The enhanced interactivity offered by CSR frameworks like React.js in regions with robust network connectivity can result in significantly longer loading times in areas with limited bandwidth \cite{nordstrom2023comparisons, Iskandar2020CSRvsSSR}. This disparity in performance can disproportionately affect users in regions with poor network infrastructure. This discrepancy arises because CSR frameworks transfer the entirety of the webpage data to the client side for processing~\cite{halfNineServerSideClientSide, Iskandar2020CSRvsSSR}, necessitating users in low-bandwidth environments to download substantial amounts of data and render the application on devices with limited processing.
 
Koradia et al.~\cite{KoradiaCellularIndia} reported cellular data connectivity latencies in India reaching up to 1200 ms. When compounded by the data processing demands of CSR on the client side, such latencies can significantly amplify delays and diminish user experience.

Additionally, in bandwidth-constrained environments, TCP flows encounter high packet loss rates, severe unfairness, and repetitive timeouts, which could render CSR-based websites practically unusable~\cite{Zaki2014WebLatencyGhana}. Furthermore, by requiring low-end devices to process and render entire applications locally after downloading substantial data, CSR frameworks place a disproportionate burden on users in these regions, leading to poor performance and significantly hampered access to the digital economy. This challenge exacerbates the digital divide, as individuals with limited internet access and processing capabilities experience slower and less responsive websites, hindering their ability to  participate in the digital economy \cite{grigorik2013high, halfNineServerSideClientSide}.

One of the main objectives behind this work is to ensure that the adoption of novel technologies like Next.js, particularly at scale, is conducted responsibly, promoting equitable access to digital resources. With features like SSR, static site generation (SSG), and caching by default, Next.js seems like a promising solution to reduce client-side processing and improve performance for users in low-resource settings. However, it is essential to validate the efficacy of these assumptions through rigorous evaluation to ensure that its adoption addresses the challenges faced in developing regions.
\vspace{-10pt}
\subsection{Businesses and other Institutions}

In 2006, Amazon showed that each 100ms increase in page load time correlated with a 1\% reduction in sales, representing an annual loss of \$107 million---an impact equivalent to \$3.8 billion~\cite{AmazonPageSpeedStudy}.

This study shows how critical web performance is to business success. However, the reliance on CSR introduces significant drawbacks, as the need for browsers to download and execute large JavaScript files can lead to slower initial page loads~\cite{muzeel}, frustrating users and increasing bounce rates, particularly among those with older or less powerful devices.  With studies showing that a 100ms delay can reduce conversion rates by 7\% and 53\% of mobile users abandoning sites that take over three seconds to load~\cite{AkamaiPerformanceReport2017}, it becomes imperative for companies adopting newer technologies like Next.js to ensure these migrations deliver measurable benefits especially in terms of faster performance and better search engine indexing ~\cite{halfNineServerSideClientSide}. Next.js holds the potential to drive better user experiences, but once again, validating these claims is essential to justify the significant investment in such transitions.
\vspace{-10pt}
\section{Evaluation Methodology}
The methodology of this work is designed in two parts: a) a quantitative metrics analysis and b) a qualitative user experience survey. Next we explain the approach taken by each part. 

\subsection{Quantitative Metrics Analysis}
In the first part of the analysis, we compared important performance metrics~\cite{WebPerformanceMetrics} relating to the different timing stages of loading a page: metrics such as First Contentful Paint (FCP), Time to Interactive (TTI), Page Load Time (PLT), and Lighthouse SEO scores of websites built using both React.js and Next.js.

One of the main challenges that we faced was to find a dataset that contains lists of websites that are built using both React.js and Next.js. While some organizations may have migrated from one framework to the other, it is uncommon for both versions to remain operational and receive consistent maintenance. We had access to a large number of Next.js-only or React.js-only websites, but it was obvious that comparing unrelated websites would not provide meaningful insights. So to ensure a fair comparison, we had to create two sets of the same websites using both React.js and Next.js, which was especially challenging because Next.js uses a file-system based router~\cite{nextjs_layouts_pages}.

\subsubsection{Defining the Testing Framework}
First, we began by outlining the specific features offered by Next.js by default to determine which types of websites would be the most meaningful to create. The features that were centered around performance optimization were partial re-rendering, pre-fetching, soft navigation, code splitting, caching, parallel routes, intercepting routes, server actions, and incremental static regeneration \cite{pati2024nextjs}. On top of these, the element-level optimizations for images, videos, fonts, etc. could be additionally tested in any of the websites created. 

As explained earlier, since the only way to obtain a dataset of equivalent websites was to create it ourselves, we made sure to rigorously choose a diverse set of applications, where some would target specific features, while others would test all-around performance, ensuring a comprehensive evaluation of real-world use cases. Seven pairs of websites and three pairs of web apps (with identical back-ends) were carefully chosen for this purpose. These included a travel destination explorer, a to-do app, a Wikipedia article viewer, a real-time stock tracker, a GitHub project portfolio, a travel booking website, etc \cite{pati2024nextjs}. Many of these had additional integrations like embedded maps (leaflet), charting libraries like Recharts, styling libraries like Tailwind and Bootstrap, APIs capable of sourcing real-time stock analytics, GitHub projects, Wikipedia articles, etc \cite{pati2024nextjs}. Two of the web apps were full-scale clones of popular applications---Amazon and YouTube. This enabled us to perform a more holistic assessment by testing the two frameworks under different conditions. 

\begin{figure*}[htb] 
\centering 
    \subfigure[Without CPU throttling.]{ \begin{minipage}[b]{0.4\linewidth} \centering \includegraphics[width=2.1in]{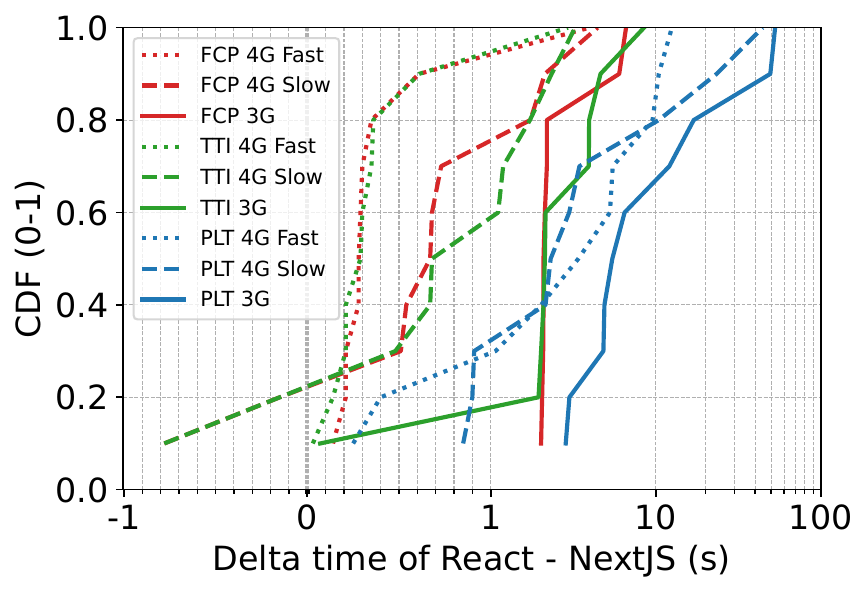} \label{fig:quantitative_a} \end{minipage} }
    \subfigure[With CPU throttling.]{ \begin{minipage}[b]{0.4\linewidth} \centering \includegraphics[width=2.1in]{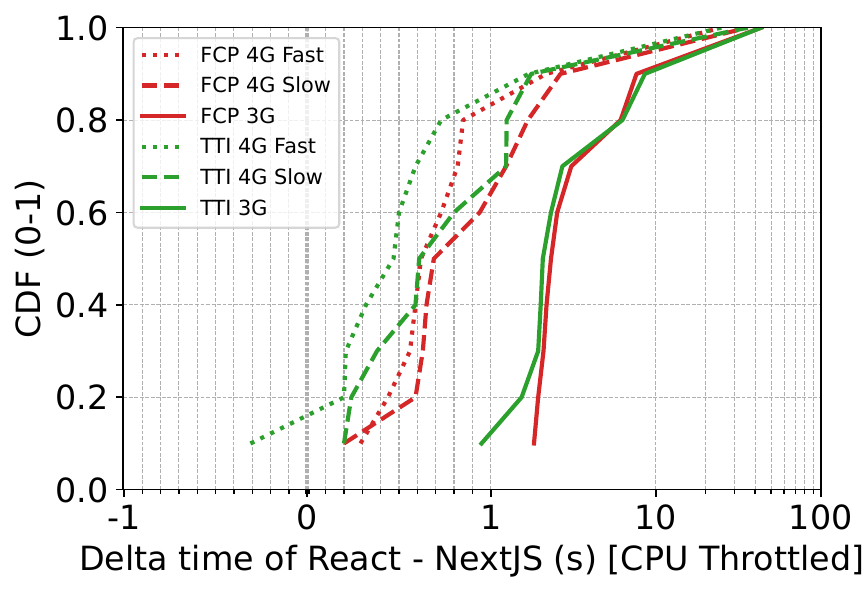} \label{fig:quantitative_b} \end{minipage} }
    \vspace{-0.15in}
\caption{Quantitative evaluation results of Next.js vs.\ Ract across different network conditions.} 
\vspace{-0.15in}
\label{fig:quantitative}
\end{figure*}

\subsubsection{LLM-Assisted Migration Workflow}
To make the process of creating the websites and web apps simpler, we decided on implementing an LLM-assisted migration workflow that involved first creating the websites on React.js and then using Claude-engineer to set up and migrate to a Next.js project. Claude-engineer~\cite{claudeengineer2024}, an advanced interactive command-line interface, utilizes the capabilities of the Claude 3.5 Sonnet LLM in conjunction with local system functions to access files, initiate projects, and execute code. It was well-trained on the Next.js documentation along with detailed migration instructions recommended by Next.js~\cite{nextjs_vite_migration}.

Although the initial draft of a Next.js project was generated using the LLM-assisted workflow, subsequent manual adjustments were necessary to fully leverage Next.js's optimization features. This refinement process required a significant time investment. However, we realized that such a flow of using the semi-automated LLM-assisted migration workflow could go on to become the recommended workflow for similar migrations. The final versions of the websites were hosted on Vercel \cite{vercel_home}.

\subsubsection{Performance Testing and Analysis}
 
For the purpose of performance testing, we decided on using FCP and TTI in three different network conditions: 4G Fast  (10 Mbps, 40ms RTT), 4G Slow (9 Mbps, 170ms RTT), and 3G (1.6 Mbps/768 Kbps 300ms RTT), under both normal and 6x lower CPU-throttled conditions to measure the initial page load metrics using the ``Performance Window'' in the Chrome Developer Tools~\cite{chrome_devtools_performance_reference}. Metrics such as FCP and TTI are vital as they directly reflect the user's real-world experience of loading and interactivity ~\cite{WebPerformanceMetrics}. We also measured the lighthouse SEOs~\cite{chrome_lighthouse}. 

To extensively evaluate the websites, we simulated identical sets of actions - for e.g, clicking something, searching something, navigating, etc. -  on both versions and recorded the `Finish' time in the network tab to determine the final loading time~\cite{chrome_devtools_network_reference}. The `Finish' time was unaffected by manual delays during the simulations, as it exclusively measured the total time taken for loading all the resources requested. However, due to limitations of the Network tab, we could not replicate these tests under throttled CPU conditions.
Furthermore, the network tab was paused at the end to prevent any more network requests before reading the `Finish' time. This testing methodology was replicated across all three network conditions. However, inherent limitations of the network tab precluded the simulation of the throttled CPU conditions. To prevent any selector's bias in sourcing these values, we made sure to take the median of three runs for every single metric across all the websites.

\vspace{-7pt}
\subsection{User-Centric Qualitative Assessment}
In addition to the quantitative evaluation, we also incorporated a qualitative evaluation to gather user feedback on the usability and performance of the websites. This was achieved through participatory research studies, where a diverse group of participants from NYU Abu Dhabi were recruited. 67 participants with devices of varying specifications of GPU, RAM, and other hardware configurations were included to analyze the website performance under real-world contexts. They provided feedback based on their experiences with loading times, interaction response times, navigation smoothness, and the overall browsing experience.

The diversity of the NYU Abu Dhabi community, with students from over 100 countries studying across several disciplines, was chosen to bring varied perspectives and responses to the study. 
Two sets of surveys, where the ordering of the websites was changed to prevent bias, were sent out to the participants. The participants evaluated two full-scale clones of Amazon and YouTube created in both Next.js and React. While only two sets of web applications were tested, this decision was deliberate and necessary to ensure that the study could focus on the overall quality of user interaction in a robust and controlled manner. The selection of these applications was predicated on their capacity to rigorously evaluate the capabilities of Next.js, encompassing a diverse set of features representative of essential web development use cases. By selecting clones of globally significant applications with varying use cases and design approaches, we ensured that the insights drawn would not be skewed by edge cases allowing us to source statistically significant data and derive meaningful conclusions, ensuring the validity and impact of the study. 

We made sure that the links and questions were presented as ``Version 1'' and ``Version 2'' to prevent bias based on prior understanding of the technologies. We also ensured that participants understood this study was not focused on the interface design or visual aesthetics but rather on evaluating the overall user experience, emphasizing speed and smoothness during website navigation.

\vspace{-10pt}
\section{Results}

\begin{figure*}[htb] 
\centering 
    \subfigure[YouTube - Faster Navigation]{ \begin{minipage}[b]{0.3\textwidth} \centering \includegraphics[width=2.2in]{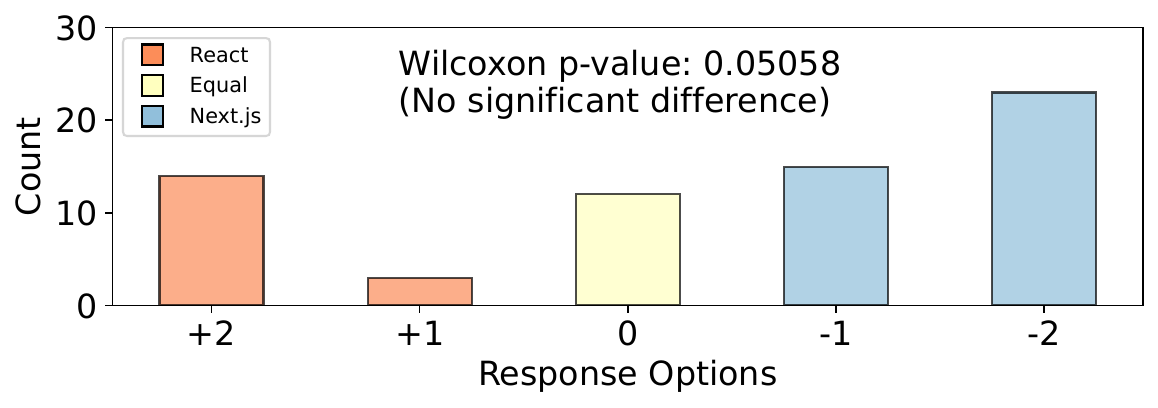} \label{fig:youtube_a} \end{minipage} }\quad
    \subfigure[YouTube - Loading And Interaction-Response]{ \begin{minipage}[b]{0.3\textwidth} \centering \includegraphics[width=2.2in]{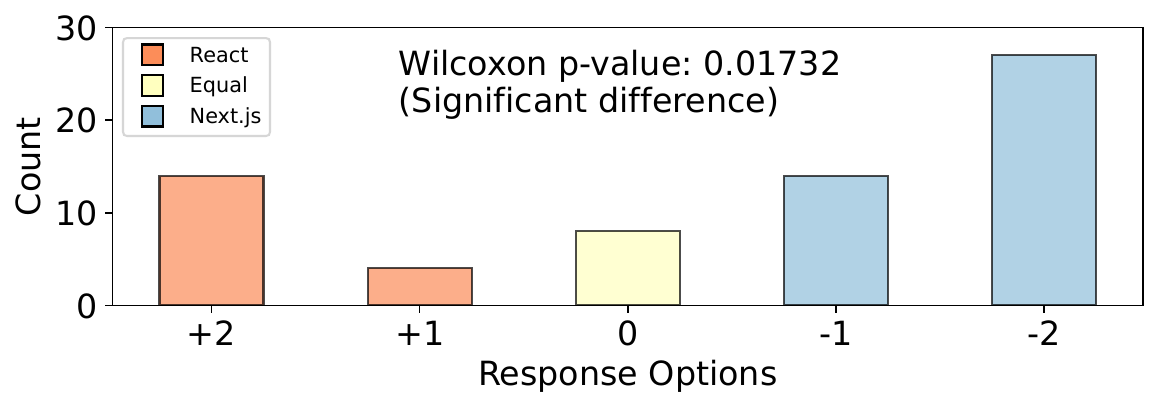} \label{fig:youtube_b} \end{minipage} }~
    \quad
    \subfigure[YouTube - Smoother Browsing Experience]{ \begin{minipage}[b]{0.3\textwidth} \centering \includegraphics[width=2.2in]{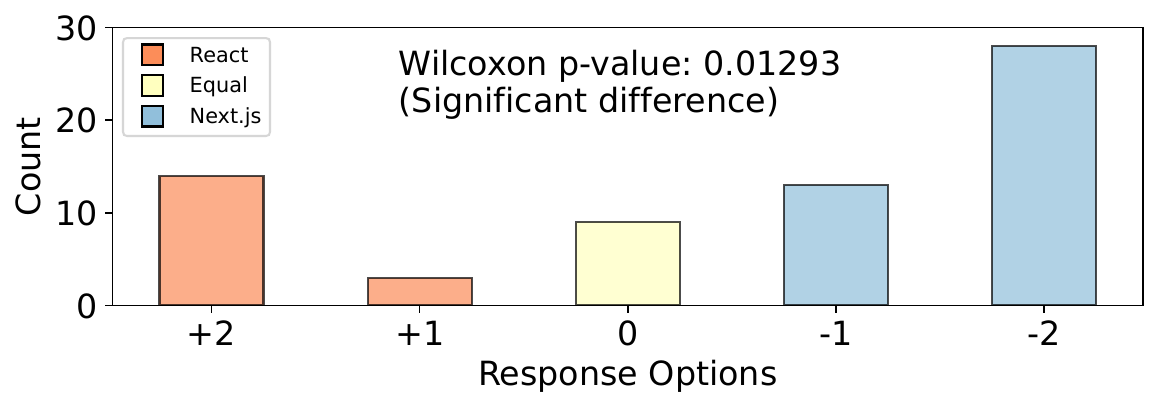} \label{fig:youtube_c} \end{minipage} } \\
    \subfigure[Amazon - Faster Navigation]{ \begin{minipage}[b]{0.3\textwidth} \centering \includegraphics[width=2.2in]{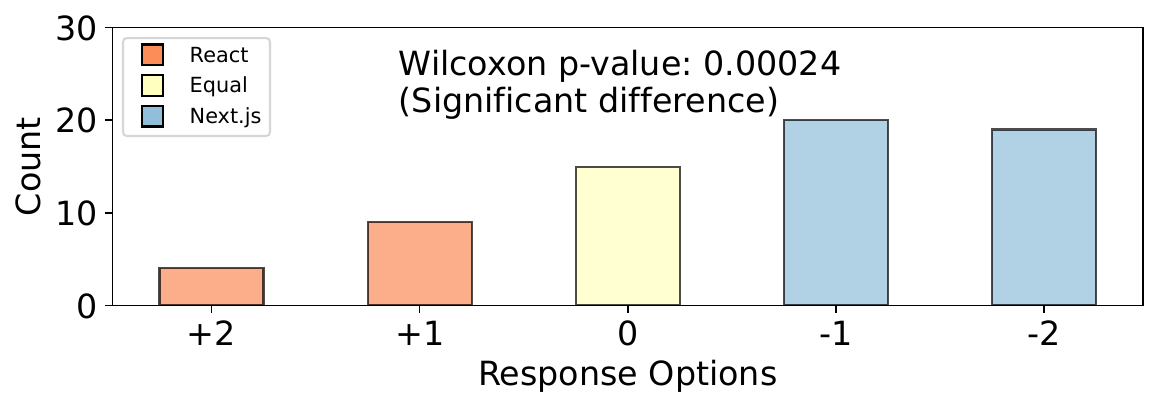} \label{fig:amazon_a} \end{minipage} }\quad
    \subfigure[Amazon - Loading And Interaction-Response]{ \begin{minipage}[b]{0.3\textwidth} \centering \includegraphics[width=2.2in]{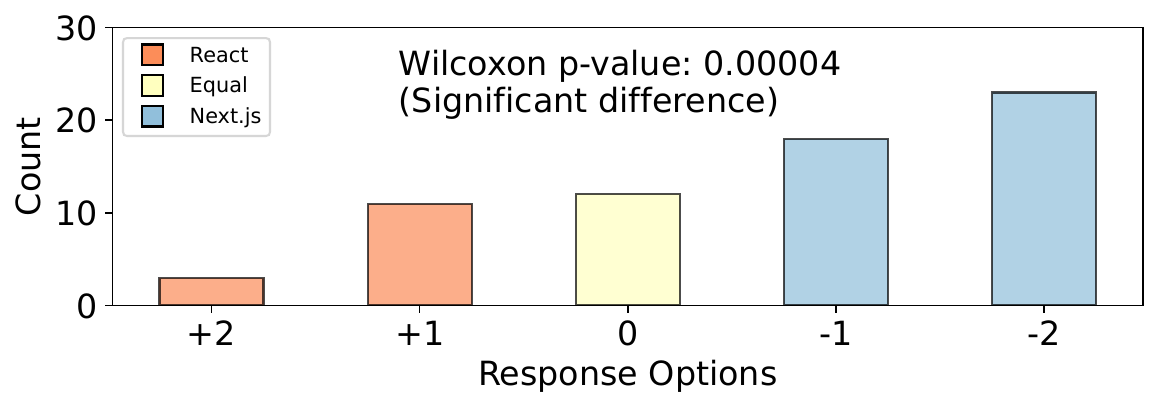} \label{fig:amazon_b} \end{minipage} }
    \quad
    \subfigure[Amazon - Smoother Browsing Experience]{ \begin{minipage}[b]{0.3\textwidth} \centering \includegraphics[width=2.2in]{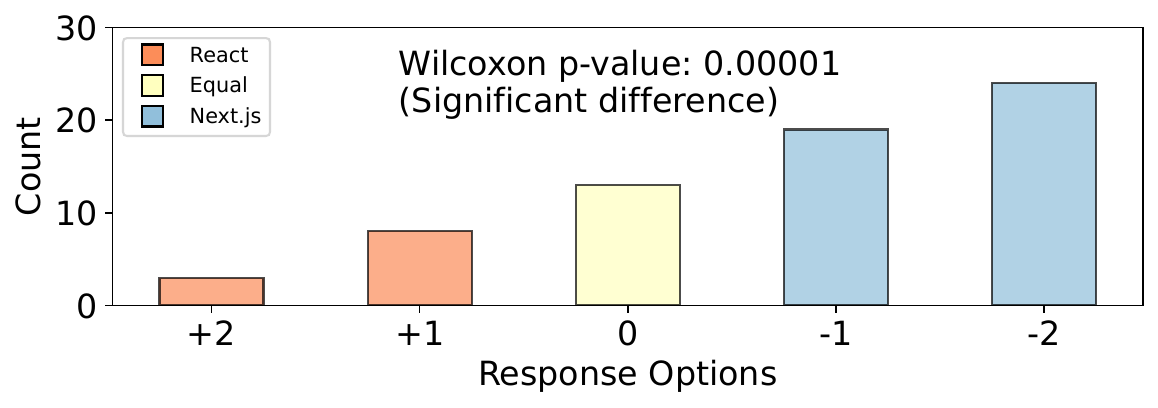} \label{fig:amazon_c} \end{minipage} } 
    \vspace{-0.15in}
\caption{YouTube and Amazon clone websites quantitative evaluation.} 
\vspace{-0.15in}
\label{fig:qualitative}
\end{figure*}

\subsection{Quantitative Results}

After obtaining the data points for each metric across all conditions, we calculated the delta time between the React and Next.js versions of each website for every metric. This delta time was computed as the difference between the corresponding values of React and Next.js (React - Next.js), providing a direct comparison of performance. Figures~\ref{fig:quantitative_a} and~\ref{fig:quantitative_b} show the cumulative distribution functions (CDFs) for each metric under all network conditions (fast 4G, slow 4G, and 3G) for both the normal CPU and throttled conditions, respectively. The results show that the delta values are positive for almost all metrics and network conditions, apart from a few outliers in the case of the TTI metric for fast and slow 4G (dotted and dashed green curves). Additionally, the results also highlight that with slower network connectivity, all timing metrics are improved in webpages built with Next.js compared to React, as well as with throttled CPUs. This highlights the significance of using Next.js to improve the overall performance for users in developing regions suffering from both poorer network connectivity and over-reliance on low-end mobile devices. Finally, for the SEO evaluations, Next.js received an average score of 100\% compared to React, which received a score of 88.8\%. These results therefore help us conclusively answer both \textbf{RQ1} - that Next.js indeed has overall faster performance compared to React.js across different network conditions and device processing capabilities - and \textbf{RQ2} - that Next.js has better SEO metrics than React.js.

\vspace{-10pt}
\subsection{Qualitative Survey Insights}
The survey received a total of 67 responses, representing approximately 30 distinct device configurations. For every question, the participants had an option to choose on a scale of 5 values: ``Next.js/React.js was clearly better,'' ``Next.js/React.js was slightly better,'' and ``Both are equivalent.'' We encoded these to integer values from -2 to +2, taking 0 (equivalent) as our null hypothesis. Then using the Wilcoxon p-value method, we ascertained which metrics showed statistically significant results.

Figure~\ref{fig:qualitative} shows the response results, highlighting  that participants preferred the Next.js version of the YouTube clone for loading and interaction-response times and the overall smoother browsing experience. The survey also showed that participants preferred the Next.js version of the Amazon clone for all three metrics: faster navigation, loading and interaction-response times, and a smoother browsing experience. This provides an answer to \textbf{RQ3}, where Next.js does offer a better user experience than React.js.

\vspace{-8pt}

\section{Conclusion}
Our results, derived from both quantitative metrics and qualitative user feedback, provide strong evidence that Next.js significantly improves the performance metrics, especially for poor network conditions, while maintaining an improved user interaction experience. Beyond these findings, the research highlighted the importance of equitable access to web resources, emphasizing Next.js's role in mitigating the digital divide and fostering global network equity. This study lays the foundation for broader investigations into the scalability and adaptability of Next.js (and other hybrid frameworks) across a wider spectrum of use cases, including enterprise-level applications and globally distributed user bases. By expanding the scope to include larger datasets and more complex user scenarios, future research can deepen our understanding of Next.js's potential in shaping a more inclusive and high-performing digital ecosystem. 

\bibliographystyle{ACM-Reference-Format}
\bibliography{sample-base}

\end{document}